\documentstyle[11pt]{article}

\newcommand{\be}{\begin{equation}}
\newcommand{\ee}{\end{equation}}
\newcommand{\bea}{\begin{eqnarray*}}
\newcommand{\eea}{\end{eqnarray*}}
\newcommand{\ba}{\begin{eqnarray}}
\newcommand{\ea}{\end{eqnarray}}
\renewcommand{\baselinestretch}{2}
\newcommand{\id}
{1 \hspace{-1mm}\raisebox{0.8mm}{$\scriptstyle |$}}

\begin{document}
\begin{flushright} UCL-IPT-99-06
\end{flushright}

\vspace*{5mm}

\begin{center}
\Large{\bf Phases and Amplitudes in Inclusive $\Psi$ and $\Psi'$ Decays}
\end{center}

\vspace*{15mm}

\begin{center}
\Large{\it J.-M. G\'erard and J. Weyers }
\end{center}

\vspace*{5mm}

\begin{center}
Institut de Physique Th\'eorique\\
Universit\'e catholique de Louvain \\
 B-1348 Louvain-la-Neuve,
Belgium
\end{center}

\thispagestyle{empty}

\vspace*{40mm}

\renewcommand{\baselinestretch}{1.4}

\begin{abstract}

In inclusive decays of the $\Psi$ (3097), electromagnetic and gluonic
annihilation
amplitudes add incoherently, namely they are 90$^\circ$ out of phase. We
argue that
this incoherence must persist in each exclusive decay channel. For
inclusive $\Psi'$
(3686) decays, we suggest the absence of a significant direct annihilation
amplitude
into three gluons and propose a new amplitude via QCD anomalies and the
$h_c$ (3526)
off shell. Phenomenological implications for exclusive decay channels are
pointed out.

\end{abstract}

\renewcommand{\baselinestretch}{2}

\newpage

\section{Introduction}

>From a theoretical point of view, two generic classes of transition
amplitudes have
to be considered in order to describe $\Psi$ (3097) and $\Psi'$ (3686) decays :
survival amplitudes and annihilation amplitudes.

By survival amplitudes (on shell) we mean transition amplitudes from the
$\Psi$ or
$\Psi'$ to a final state which contains a bound $(c\bar c)$ pair. Of course,
annihilation amplitudes correspond to transitions from $\Psi$ or $\Psi'$ to
quarkless
intermediate states.

In the context of QCD, survival and annihilation amplitudes are radically
different :
while the latter can be reasonably handled as perturbative processes, the
former are
intrinsically non perturbative. Because of the scale and axial anomalies,
the largest
survival amplitude is expected \cite{1} to correspond to a transition from
the $\Psi'$
to the
$\Psi$ accompanied with gluons in a $0^{++}$ or $0^{-+}$ state which then
hadronize.
Needless to say, this agrees with the data. The
dominant role of survival amplitudes is also illustrated by the large decay
rates for
radiative processes such as $\Psi \to \eta_c \gamma$ and $\Psi' \to \chi
\gamma$.

Since the $\Psi$ is the (next to) lowest state in the charmonium spectrum, the
radiative decay $\Psi \to \eta_c \gamma$ exhausts the possible
contributions from
survival amplitudes. To leading order in QCD and QED, one is then left with
perturbative annihilation amplitudes of the $\Psi$ into three gluons or
into one
photon. One of the purposes of this note is to draw attention on the
relative phase
of these annihilation amplitudes. This will be done in the next paragraph.
The main
result is that we expect a {\em universal incoherence} of these amplitudes,
that is to
say they are 90$^\circ$ out of phase in every exclusive decay channel. This
solves
the ``phase problem" in $\Psi$ decays, at least in the approximation where
bona fide final state interactions are neglected.

For the $\Psi'$, on the other hand, the overall phenomenological picture is
much less
clear. While there is no doubt that survival amplitudes account for the bulk of
$\Psi'$ decays, the important question is to correctly identify the leading
strong
decay amplitude which is responsible for the remaining 20$\%$.
Usually this amplitude is assumed to be a direct annihilation of the
$\Psi'$ into
three gluons. In
$\S$ 3 we will argue first of all that there is overwhelming phenomenological
evidence against this scenario. We then propose a \underline{new} amplitude
for the
strong annihilation of the $\Psi'$ into light hadrons namely an off shell
survival
amplitude where the $c \bar c$ pair has the quantum numbers of the not yet
well-established
$h_c$ (3526) i.e.
$J^{PC} = 1^{+-}$. More precisely, we suggest that the strong annihilation
of the
$\Psi'$ into light hadrons is a two step process : in the first step the
$\Psi'$
goes, via anomalies, into two gluons in a $0^{++}$ or $0^{-+}$ state and an
off shell
$h_c$ (3526); in the second step the off shell $h_c$ annihilates into three
gluons
to produce light hadrons. In spirit, our model is somewhat
akin to the Gell-Mann, Sharp and Wagner model \cite{2} for the decay
$\omega \to 3\pi$
: here, the dominant intermediate state is an off-shell $\rho$ accompanied by a
$\pi$. There are many possible tests of the model we advocate, some of
which will be
mentioned at the end of $\S$3.

Finally, to conclude this note, we briefly point out, in paragraph 4, some
obvious
but important phenomenological implications of our model for exclusive
$\Psi'$ decay
modes. In particular, there is no direct strong annihilation amplitude for
$\Psi'
\to
\rho
\pi$ i.e. the ``$\rho \pi$ puzzle" \cite{3} is solved in our model : the
so-called
14$\%$ rule between $\Psi'$ and $\Psi$ branching ratios is expected to be valid
\underline{only} for purely electromagnetic annihilation processes.

\newpage

\section{Phases in inclusive and exclusive $\Psi$ decays}

In the conventional picture of inclusive hadronic $\Psi$ decays, which we
adopt, the
amplitude for  $\Psi$ going into light hadrons is given by
\be
A (\Psi \to \ \mbox{hadrons}) \equiv A^H_g + A^H_\gamma.
\ee
$A^H_g$ is the hadronized (three gluon) QCD annihilation amplitude and,
similarly,
$A^H_\gamma$ is the hadronized (one photon) QED annihilation amplitude. A
priori,
$A^H_g$  and
$A^H_\gamma$ are complex numbers.

Experimentally \cite{4}
\ba
Br \ (\Psi \to \mbox{hadrons}) &=& 1 - Br (\Psi \to \eta_c \gamma, \ell^+
\ell^-)\\
&\approx& 86.7 \% \nonumber
\ea
but notwithstanding Eq.(1) one usually writes Eq.(2) in the form
\be
Br \ (\Psi \to \mbox{hadrons}) = B (\Psi \to \mbox{gluons} \to
\mbox{hadrons}) + B
(\Psi \to \mbox{photon} \to \mbox{hadrons}).
\ee

Clearly this last equation holds \underline{only} if there is \underline{no
interference} between $A^H_g$  and $A^H_\gamma$ or in other words, only if
these
amplitudes are 90$^\circ$ out of phase and thus add incoherently.

If $\varphi$ is the relative phase angle between $A^H_g$  and $A^H_\gamma$,
the data
per se, i.e. Eq.(2) together with
\ba
Br (\Psi \to \mbox{photon}  \to  \mbox{hadrons})   &\approx &  3
\sum Q^2_i (1 +
\frac{\alpha_s}{\pi}) Br (\Psi \to \mu^+ \mu^-)
\\
& \approx & 13\%
\nonumber
\ea
put only a mild constraint on its value
\be
\varphi {<\hspace{-3.5mm}\raisebox{-1mm}{$\scriptstyle \sim$}} \  110^\circ.
\ee

There are, however, strong theoretical arguments in favor of $\varphi = \pi/2$.
Indeed this value follows directly from the orthogonality of the three
gluon and
virtual photon states to leading order. In a symbolic but obvious notation,
one has
\be
A^H_g = \sum_h \langle h | 3g \rangle \langle 3g | \Psi \rangle
\ee
and
\be
A^H_\gamma = \sum_h \langle h | \gamma \rangle \langle \gamma | \Psi \rangle
\ee
Then, clearly,
\be
A^{\ast H}_g A^H_\gamma = \langle \Psi | 3g \rangle \langle 3g | (\sum_h |
h \rangle
\langle h | ) | \gamma \rangle \langle \gamma | \Psi\rangle = 0
\ee
is equivalent to
\be
\langle 3g | \gamma \rangle = 0
\ee
since $\sum_h | h \rangle \langle h | = \id$.

Incoherence between $A^H_g$ and $A^H_\gamma$, or Eq.(3), has thus nothing
to do with
the hadronization process nor with the final states : it simply follows
from the
orthogonality relation, Eq.(9).

Incoherence or non interference at the inclusive level implies either non
interference
in every single exclusive channel or a conspiracy between channels. The latter
possibility appears to be ruled out. Indeed consider all amplitudes as
functions of
$m_\Psi$ (or $m_c$). Varying this parameter does not affect annihilation nor
hadronization except in trivial ways i.e. it opens up or closes down, one
at the
time, a possible exclusive channel depending on its threshold. Thus, each
of these
channels must, by itself, exhibit non interference and hence we expect the
latter
property to hold channel by channel.

By this simple argument we thus expect {\em universal incoherence},
exclusive channel
by exclusive channel, between the QCD and QED annihilation amplitudes of
$\Psi$.

It has been known for quite a while that there is a very large phase angle
of the
order of $\frac{\pi}{2}$ between the electromagnetic and the gluonic decay
amplitudes
of the $\Psi$ into two pseudoscalars as well as into a pseudoscalar and a
vector
\cite{5} or into a nucleon-antinucleon pair \cite{5,6}. This was recently
rediscovered \cite{7}, at least in the mesonic channels, and interpreted as
a large
``final state interaction" phase \cite{7,8}.

The arguments presented above prove that the phase angle of $\frac{\pi}{2}$
is {\em
independent} of the final states but originates from non interacting or
orthogonal
intermediate states.

To conclude this section let us add a comment and a caveat. The comment is
that,
despite large errors in the branching ratios, the data on $\Psi \to$ tensor
meson +
vector meson also appear compatible with non interference thus
strenghtening our
conclusion. The caveat is that in extracting the relative phase between
$A_g$ and
$A_\gamma$ from exclusive channels one should \underline{not} a priori
neglect the
genuine final state interaction phases namely the eigenphases of the
hadronic $S$
matrix. For example in the decay $\Psi \to p \bar p$, a relative phase
between the
isospin amplitudes
$A_{g,\gamma} (I=0)$ and $A_\gamma (I = 1)$ is in principle present.

\section{A new amplitude for hadronic $\Psi'$ decays}

As mentioned in the introduction, survival amplitudes are responsible for
the bulk of
$\Psi'$ decays. However, the $\Psi'$ does also decay into light hadrons
\cite{4}
\ba
Br \  (\Psi' \to \mbox{hadrons}) &=& 1 - Br (\Psi' \to \Psi \pi\pi,  \Psi
\eta, \chi
\gamma, \eta_c \gamma, \ell^+ \ell^-) \\
&\cong& 20\%.
\nonumber
\ea

The standard description of these decays is in terms of the QCD and QED
perturbative
annihilation amplitudes $A (\Psi' \to 3g)$ and $A(\Psi' \to \gamma)$. In
our opinion,
this picture is essentially incorrect.

Indeed, let us for a moment concentrate on the three
gluon intermediate state. This is the same intermediate state as in $\Psi
\to 3g$
except that the gluons are slightly more energetic. Of course this will
open up new
hadronic  channels but for exclusive channels common to $\Psi$ and $\Psi'$
there can
{not} be significant differences in relative abundances if the three gluon
intermediate state makes any physical sense. The data and in particular the
recent
BES data blatantly contradict this theoretical expectation. The pattern of
observed
channels in $\Psi'$ decays is totally different from the one in $\Psi$
decays : no
$(\rho \pi)$ neither a $(\rho a_2)$ channel in $\Psi'$ decays while these
channels
are among the dominant ones in $\Psi$ decays. On the other hand, the $b_1 \pi$
channel appears dominant in $\Psi'$ decays while in $\Psi$ decays it is one
of many
with branching ratios of a few times $10^{-3}$. Further evidence comes from the
comparison of the ``$K^\ast - \overline K$" and ``$K_1 - \overline K$" type
channels
which exhibit different sensitivity to flavor $SU(2)$ and $SU(3)$ symmetry
breakings,
respectively. These completely different ordering patterns in
$\Psi$ and $\Psi'$ decays into two mesons are {\em incompatible} with
hadronization of
{\em identical} (in quantum numbers) intermediate states.  We believe that a
natural explanation of these facts is simply that the $\Psi'$ does not
significantly
annihilate into three gluons i.e.
\be
Br (\Psi' \to 3g) \ll 20\%.
\ee

Actually this phenomenological conclusion can also be motivated by analogy
with the
positronium data. There, the decay width of the $2^3 S_1$ state into three
photons is
a factor eight smaller than for the $1 ^3 S_1$ ! From a more theoretical
point of
view,  if both the physical
$\Psi$ and
$\Psi'$ did dominantly annihilate into three gluons, they would
\underline{mix} and could thus not be the physical  eigenstates of the
effective strong $(c \bar c)$ hamiltonian which they are. In a non relativistic
model, for example, the $\Psi'$ is simply a radial excitation of the
$\Psi$. This is a
well defined picture in which $\Psi$ and $\Psi'$ are orthogonal states. If the
annihilation into three gluons could be treated as a ``perturbation" to the non
relativistic potential, then clearly the unperturbed states would mix and
rearrange
themselves into {\em orthogonal} physical states i.e. the analog of Eq.(11)
would
obviously be true. From that point of view, it is also illuminating that the
asymptotic behavior of the ($e^+ e^- \to$ light quarks) cross-section
requires an
infinite tower of vector meson radial excitations... with vanishingly small
hadronic
decay widths.

Whatever the merit of these arguments, if Eq.(11) is phenomenologically
correct, the
obvious question is then how does the $\Psi'$ eventually annihilate into light
hadrons?

The answer is almost self evident if one goes back to the dominant
transitions observed in the
$c \bar c$ system, namely the transition from the $\Psi' (1^{--})$ to the $\Psi
(1^{--})$ via the scale or axial anomaly, i.e. gluons in a $0^{++}$ or
$0^{-+}$ state
\cite{1} :
\be
\Psi' \to \Psi (1^{--}) + (0^{++} \ \mbox{or} \ 0^{-+}).
\ee

Exactly the same mechanism allows for one and \underline{only one} other
transition,
\be
\Psi' \to h_c (1^{+-}) + (0^{++} \ \mbox{or} \ 0^{-+}).
\ee
While the $\Psi$ in Eq.(12) is on shell when the two gluons hadronize into
$\pi\pi$ or
$\eta$, the
$h_c$ in Eq.(13) cannot be on shell and, as such, it has only one way to go
namely
annihilate into three gluons.

Thus the new amplitude we propose as the dominant mechanism for (light)
hadronic
decays of the $\Psi'$ corresponds to the two step process
\be
\Psi' \to h_c (1^{+-}) +  (0^{++} \ \mbox{or} \ 0^{-+}) \to 3g (1^{+-}) + 2g
(0^{++}  \ \mbox{or} \ 0^{-+}).
\ee

Specifically, our new annihilation amplitude is, in some sense, when the
charmed
quarks have finally disappeared, a {\em particular} five gluon
configuration albeit
\underline{not} a perturbative one : two of the gluons (the non
perturbative ones)
carry dominantly the quantum numbers $0^{++}$ or $0^{-+}$ while the three
others (perturbative gluons) carry the quantum numbers of the $h_c$ namely
$1^{+-}$.
To leading order, hadronization should preserve these quantum numbers.

Many possible tests of our model come to mind. Perhaps the simplest ones
are in the
analysis of inclusive spectra $\Psi' \to \pi^+ \pi^- X, \Psi' \to \eta
\widetilde
X$,  $\Psi' \to \eta' \widetilde{\widetilde X}$ : the states $X, \widetilde X,
\widetilde{\widetilde X}$ should be dominantly in a $1^{+-}$ configuration.

As far as exclusive channels are concerned, we expect $(\pi\pi) h_1 (1170)$
or $\eta
(\eta') h_1 (1170)$ to be important decay modes of the $\Psi'$ but, of
course, final
state interactions will lead to other configurations as well : for example
$SU(3)$
elastic final state interactions do allow the transition $\eta h_1 (1170)
\to \pi b_1
(1235)$ and the importance of the latter channel in the BES data \cite{9}
is perhaps
a hint that we may not be on the wrong track.

\section{Summary and Conclusions }

The main points of this note are the following
\begin{enumerate}
\item[-] in $\Psi$ decays into light hadrons, the three gluon annihilation
amplitude
and the QED amplitude add incoherently in all channels ;
\item[-] in $\Psi'$ decays into light hadrons, the dominant QCD annihilation
amplitude is \underline{not} into three gluons but via a two step process
into a
specific configuration of five gluons.
\end{enumerate}

To conclude let us simply point out some obvious phenomenological
consequences of all
our qualitative arguments. Our model for $\Psi'$ hadronic decays predicts a
sizeable
$\Psi' \to (\pi^+ \pi^-$ or $\eta) X (1^{+-})$ branching ratio. At the
exclusive
level, it implies that :
\begin{enumerate}
\item[a)] to leading order there is \underline{no} strong decay amplitude
for the
processes $\Psi' \to \rho \pi$ and $\Psi' \to K^\ast \overline K$;
\item[b)] the well-known 14$\%$ rule relating $\Psi'$ and $\Psi$ branching
ratios
should hold for hadronic processes like $\Psi^{(')} \to \omega \pi^0$ which
take place
via the QED amplitude
\underline{only} \cite{10}.
\end{enumerate}
Finally we note that our model can easily be extended to the $\Upsilon$ system.

\newpage

\par\noindent
{\Large{\bf Acknowledgements}}

We have enjoyed many conversations with J. Pestieau and C. Smith.

\end{document}